\documentclass{jpconf}

\usepackage{amssymb}
\usepackage{graphicx}

\newcommand{\be}{\begin{equation}}
\newcommand{\ee}{\end{equation}}
\newcommand{\bra}{\langle}
\newcommand{\ket}{\rangle}
\newcommand{\bea}{\begin{eqnarray}}
\newcommand{\eea}{\end{eqnarray}}
\newcommand{\dis}{\displaystyle}

\begin{document}

\title{Empirical Analysis of Stochastic Volatility Model by Hybrid Monte Carlo Algorithm}

\author{Tetsuya Takaishi}

\address{Hiroshima University of Economics, Hiroshima 731-0192, JAPAN}

\ead{takaishi@hiroshima-u.ac.jp}

\begin{abstract}
The stochastic volatility model is one of volatility models which 
infer latent volatility of asset returns.
The Bayesian inference of the stochastic volatility (SV) model
is performed by the hybrid  Monte Carlo (HMC) algorithm which is  
superior to other Markov Chain Monte Carlo methods in
sampling volatility variables.
We perform the HMC simulations of the SV model for 
two liquid stock returns traded on the Tokyo Stock Exchange and 
measure the volatilities of those stock returns.
Then we calculate the accuracy of the volatility measurement 
using the realized volatility as a proxy of the true volatility 
and compare the SV model with the GARCH model which is one of other volatility models. 
Using the accuracy calculated with the realized volatility 
we find that empirically the SV model performs  better than the GARCH model.

\end{abstract}


\section{Introduction}
Statistical properties of asset price returns have been extensively studied both in finance and econophysics.
It is well-known that asset price returns show  fat-tailed distributions and volatility clustering
which are now classified as stylized facts\cite{Stanley,CONT}.
Although in empirical finance the volatility is an important value to measure the risk
it is quite difficult to extract the true volatility  from asset price returns themselves. 
A popular technique to measure the volatility is to use volatility models which mimic 
volatility properties of asset returns such as volatility clustering.
A famous volatility model is the generalized  autoregressive conditional hetroskedasticity (GARCH) model\cite{ARCH,GARCH} where
the volatility variable changes deterministically depending on the past squared value of the returns and past volatilities.
There exist many extended versions of GARCH models, 
such as EGARCH\cite{EGARCH}, GJR\cite{GJR}, QGARCH\cite{QGARCH1,QGARCH2}, GARCH-RE\cite{GARCH-RE} models 
and many others, e.g. \cite{GARCH-Review,Xekalaki}, 
which are designed to increase the performance of the model. 

The stochastic volatility (SV) model\cite{SVMCMC1,SV} is another model 
which captures the properties of the volatility.  
Unlike the GARCH model,
the volatility of the SV model changes stochastically in time.
As a result the likelihood function of the SV model is given as 
a multiple integral of the volatility variables.
Such an integral in general is not analytically calculable 
and thus the determination of the parameters of the SV model 
by the maximum likelihood method becomes extremely difficult.
To overcome this difficulty 
the Markov Chain Monte Carlo (MCMC) method based on the Bayesian approach is proposed and developed\cite{SVMCMC1}.
In the MCMC method of the SV model one has to update not only the parameter variables but also the volatility ones 
under a joint probability distribution of the model parameters and the volatility variables. 
Although there exist various MCMC methods to perform the Bayesian program
the performance of the MCMC technique depends on each MCMC method.
A local update scheme is usually not effective for the SV model which has an increasing number of volatility variables
with the data size of return time series\cite{SVMCMC2}.
In order to improve the efficiency of the local update method 
the blocked scheme which updates several variables simultaneously
is proposed\cite{SVMCMC2,Watanabe}.

In this study we use the HMC algorithm\cite{HMC} which is a global algorithm
updating variables simultaneously.
This global property of the HMC algorithm could be advantageous for updating volatility variables of the SV model.
Actually it is shown that the HMC algorithm can decorrelate 
first enough the sampled data of the volatility variables\cite{ICIC2008,SVHMC}.
The HMC has been also applied  to the parameter estimation of the GARCH model\cite{HMCGARCH}.

Using the HMC algorithm we perform the Bayesian inference of the SV model for 
two stock returns (Mitsubishi Co. and Panasonic Co.) traded on the Tokyo Stock Exchange 
and measure the volatility of those stock returns.
In order to quantify the accuracy of the volatility estimated from the SV model
we calculate a loss function  using the realized volatility from the high-frequency intraday returns
as a proxy of the true volatility. The realized volatility as  a proxy of the true volatility
has been used for ranking GARCH-type models\cite{Ranking}.
In this study we also calculate the accuracy of the volatility 
from the GARCH model 
and compare the SV model and the GARCH model 
based on the criterion of the volatility accuracy.

\section {Stochastic Volatility Model}

The standard SV model\cite{SVMCMC1,SV} is defined by 
\be
y_t = \sigma_t \epsilon_t = \exp(h_t/2)\epsilon_t,
\label{eq:SV}
\ee
\be
h_t = \mu +\phi (h_{t-1} -\mu) +\eta_t,
\ee
where $h_t$ is defined by $h_t=\ln \sigma_t^2$,
and $\sigma_t$ is called volatility.
We also call $h_t$ volatility variable.
The error terms $\epsilon_t$ and $\eta_t$ are taken from independent normal distributions
$N(0,1)$ and $N(0,\sigma_\eta^2)$ respectively.

This model contains three parameters $\mu$, $\phi$ and $\sigma^2_\eta$ which have to be inferred 
so that the model could  match  the asset return data $y_t$ measured in the financial markets.
Let $\theta$ be an abbreviation for $\mu$, $\phi$ and $\sigma^2_\eta$, i.e. $\theta=( \mu, \phi,\sigma^2_\eta)$.
Then the likelihood function $L({\bf \theta})$ of the SV model can be written as 
\be
L({\bf \theta})=\int \prod_{t=1}^n f(\epsilon_t|\sigma_t^2) f(h_t|\theta) dh_1dh_2...dh_n,
\label{LFUNC}
\ee
where
\be
f(\epsilon_t|\sigma_t^2)=\left(2\pi \sigma_t^2\right)^{-\frac12}\exp\left(-\frac{y_t^2}{2\sigma_t^2}\right),
\ee
\be
f(h_1|\theta)=\left(\frac{2\pi \sigma_\eta^2}{1-\phi^2}\right)^{-\frac12}  
\exp\left(-\frac{[h_1-\mu]^2}{2\sigma_\eta^2/(1-\phi^2)}\right),
\ee

\be
f(h_t|\theta)=\left(2\pi\sigma_\eta^2\right)^{-\frac12} 
\exp\left(-\frac{[h_t-\mu-\phi(h_{t-1}-\mu)]^2 }{2\sigma_\eta^2}\right).
\ee

Unlike the GARCH model where the maximum likelihood method should work, 
$L({\bf \theta})$ of the SV model is constructed as a multiple integral of the volatility variables 
which makes the maximum likelihood estimation difficult. 
A possible estimation technique for the SV model is the Bayesian inference  performed 
by the MCMC method.
In the Bayesian inference model parameters are inferred as 
the expectation values given by
\be
\bra {\bf \theta_i} \ket = \int {\bf \theta_i} f(\theta|y) \Pi_{k=1}^3 d\theta_k,
\label{eq:INT}
\ee
where $(\theta_1,\theta_2,\theta_3)=( \mu, \phi,\sigma^2_\eta)$
and $f(\theta|y)$ is the probability distribution of $\theta$ constructed 
by the likelihood function and the prior distribution of $\pi(\theta)$ as
\be
f(\theta|y) \sim L({\bf \theta}) \pi(\bf \theta).
\ee

In general eq.(\ref{eq:INT}) can not be performed analytically and 
usually it is estimated numerically by the MCMC method.
Since $L(\theta)$ also contains the integral of volatility variables $h_t$ 
we have to update not only the model parameter $\theta$ but also the volatility variables $h_t$
in the MCMC method.
To update the model parameter $\theta$ we follow the standard update
technique\cite{SVMCMC1,SV}.

The probability distribution of the volatility variables $h_t$ is given by 
\bea
\label{eq:ham}
& P(h_t)\equiv P(h_1,h_2,...,h_n)  \sim \vspace{2cm}  \\ \nonumber
& \exp \left(-\sum_{i=1}^n \{\frac{h_i}{2}+\frac{\epsilon_i^2}{2}e^{-h_i}\}
-\frac{[h_1-\mu]^2}{2\sigma_\eta^2/(1-\phi^2)}
-\sum_{i=2}^n \frac{[h_i-\mu-\phi(h_{i-1}-\mu)]^2}{2\sigma_\eta^2}\right).
\eea
To update the volatility variables with this probability distribution we use the HMC algorithm.

\section{Hybrid Monte Carlo Algorithm}
The HMC algorithm is a global MCMC one which can update simultaneously
all the variables of the model we consider.
Originally the HMC algorithm is developed for  
the MCMC simulations of the lattice Quantum Chromo Dynamics (QCD) calculations\cite{HMC,UKAWA}
where a very large scale simulation with  a huge number of variables is inevitable.
The basic idea of  the HMC algorithm is a combination of molecular dynamics (MD) simulation 
and Metropolis accept/reject step. 
In the HMC algorithm we first introduce  momentum variables $p_i$ conjugate to $h_i$ and define
the Hamiltonian of the SV model\cite{ICIC2008,SVHMC} as
\be
H(p_t,h_t)=\sum_{i=1}^n \frac12 p_i^2 +
\sum_{i=1}^n \{\frac{h_i}{2}+\frac{\epsilon_i^2}{2}e^{-h_i}\}
+\frac{[h_1-\mu]^2}{2\sigma_\eta^2/(1-\phi^2)}
+\sum_{i=2}^n \frac{[h_i-\mu-\phi(h_{i-1}-\mu)]^2}{2\sigma_\eta^2}.
\ee
Then we integrate the Hamilton's equations of motion,
\bea
& \frac{\dis dh_i}{\dis d\tau}  = & \frac{\partial H}{\partial p_i},\\
& \frac{\dis dp_i}{\dis d\tau}  = & -\frac{\partial H}{\partial h_i},
\eea
numerically by doing the MD simulation in the fictitious time $\tau$. 
The simplest integrator for the MD simulation is 
the 2nd order leapfrog integrator\cite{HMC} given by
\bea
& h_i(\tau+\Delta \tau/2)& = h_i(\tau)+\frac{\Delta \tau}{2}p_i(\tau) \nonumber \\
& p_i(\tau+\Delta \tau)& = p_i(\tau)-\Delta \tau\frac{\partial H(p_t,h_t)}{\partial h_i} \nonumber \\
& h_i(\tau+\Delta \tau)& = h_i(\tau+\Delta \tau/2)+\frac{\Delta \tau}{2}p_i(\tau+\Delta \tau),
\label{leapfrog}
\eea
where $\Delta \tau$ is a step size of the MC simulation.
Although we adopt this  2nd order leapfrog integrator in this study 
we could also use the other improved integrators\cite{MNHOMC} 
or higher order integrators\cite{HOHMC,HOHMC2}.

After integrating the Hamilton's equations of motion up to some constant time, 
we obtain a set of new candidate variables $(p^\prime,h^\prime)$.
These candidates are accepted with the Metropolis test\cite{METRO} 
with the following probability,
\be 
P = \min\{1,\frac{\exp\left(-H(p^\prime,h^\prime)\right)}{\exp\left(-H(p,h)\right)}\}=\min\{1,\exp\left(-\Delta H\right)\},
\label{eq:ACC}
\ee
where $\Delta H$ is the energy difference given by $\Delta H = H(p^\prime,h^\prime) -H(p,h)$.
The acceptance rate of eq.(\ref{eq:ACC}) can be tuned by $\Delta \tau$. 
The high acceptance rate is not necessary for the HMC and it is shown that the optimum acceptance rate of 
the HMC with the 2nd order integrator is around 60\%\cite{HOHMC}.

\section{Realized Volatility}
In this study we use the realized volatility as a proxy of the true volatility
and use it for the accuracy calculation of the volatility inferred from the volatility models.
When high-frequency intraday return data is available 
we can construct the realized volatility  as a sum of squared intraday returns\cite{RV1,RV2,RV3}.
Let us assume that the logarithmic price process $\ln p(s)$ follows
a continuous time stochastic diffusion,
\be
d\ln p(s) =\tilde{\sigma}(s)dW(s),
\label{eq:SD}
\ee
where $W(s)$ stands for a standard Brownian motion  and $\tilde{\sigma}(s)$ is a spot volatility at time $s$.
Then the integrated volatility is defined by
\be
\sigma_T^2(t) =\int_{t}^{t+T}\tilde{\sigma}(s)^2ds,
\label{eq:int}
\ee
where $T$ stands for the interval to be integrated.
If we consider daily volatility $T$ should take one day.
Since $\tilde{\sigma}(s)$ is latent and not available from market data,
the value of eq.(\ref{eq:int}) can not be obtained directly. 

Constructing $n$ intraday returns  from high-frequency data,
the realized volatility $RV_t$ is given by a sum of squared intraday returns,
\be
RV_t =\sum_{i=1}^{n} r_{t+i\delta}^2,
\label{eq:RV}
\ee
where $\delta$ is a sampling time interval defined by $\delta=T/n$.
Note that small sampling time  interval corresponds to high sampling frequency.

If there is no microstructure noise\cite{Campbell,Zhou},  
$RV_t$ goes to the integrated volatility of eq.(\ref{eq:int}) in the limit of $n \rightarrow \infty$.
Empirically it is well-known that  the realized volatility measure suffers from the  microstructure noise
and the distortion from the  microstructure noise will be serious at very high-frequency.
Such distortion can be depicted in the volatility signature plot\cite{VSP}.  
To avoid a large distortion on the realized volatility 
and at the same time to maintain the accuracy of the realized volatility measure
we have to employ a good sampling frequency. 
The optimum sampling frequency under the independent microstructure noise was derived
and found to be between one and five minutes\cite{Bandi}

When we consider daily stock realized volatility
we have to cope with non-trading hours issue.
Usually high-frequency data are not available for the entire 24 hours.
At the Tokyo stock exchange market domestic stocks are traded in
the two trading sessions: (1)morning trading session (MS) 9:00-11:00.
(2)afternoon trading session (AS) 12:30-15:00.
The daily realized volatility calculated without including intraday returns during
the non-traded periods
can be underestimated.

An idea to circumvent the problem is
advocated by Hansen and Lunde\cite{Hansen}.
They introduced an adjustment factor which
modifies the realized volatility so that the average of the realized volatility matches
the variance of the daily returns.
Let $(R_1,...,R_N)$ be $N$ daily returns.
The adjustment factor $c$ is given by
\be
c=\frac{\sum_{t=1}^{N}(R_t-\bar{R})^2}{\sum_{t=1}^{N}RV_t},
\label{eq:HL}
\ee
where $\bar{R}$ denotes the average of $R_t$.
Here we call  $c$ the HL factor.
Then using this factor the daily realized volatility is modified to $cRV_t$.
In this study we calculate the realized volatility and the HL factor 
at several sampling frequencies and use the modified realized volatility
$cRV_t$ as a proxy of the true daily volatility.

\begin{table}
\centering
\caption{Results estimated by the HMC algorithm. {\bf SD} stands for Standard Deviation and
{\bf SE} stands for Statistical Error. The statistical errors are estimated by the jackknife method.
$\tau_{int}$ is the autocorrelation time defined by $\tau_{int} = 1 + 2\sum_{t=1}^{\infty} ACF(t)$,
where $ACF(t)$ is the autocorrelation function. We also show $h_{10}$ as a representative one of
the volatility variables $h_t$.
}
\begin{tabular}{cc|cccc}
\hline
&             & \makebox[20mm]{$\phi$}   & \makebox[20mm]{$\mu$}  &  \makebox[20mm]{$\sigma^2_{\eta}$}   &  \makebox[20mm]{$h_{10}$}    \\  \hline
Mitsubishi Co.&    &  0.987   & -7.33  &  0.0186     &  -7.48   \\
& SD          &  0.007   &  1.07  &  0.067     &  0.31   \\
& SE          &  0.0005  &  0.007 &  0.0008     & 0.01    \\
&$\tau_{int}$ &  260(130) & 1.3(1) &  800(370)  &   30(4)   \\ \hline
Panasonic Co.&   &  0.975   & -7.87  &  0.045     &  -7.57         \\
& SD          &  0.012   &  0.50  &  0.017     &   0.36  \\
& SE          &  0.0013  &  0.004 &  0.0025    &   0.01  \\
&$\tau_{int}$ & 410(250)  & 1.8(4)&  800(480)  &   35(14) \\ \hline 

\end{tabular}
\vspace{2mm}
\end{table}

\begin{figure}
\vspace{8mm}
\centering
\includegraphics[height=7.0cm]{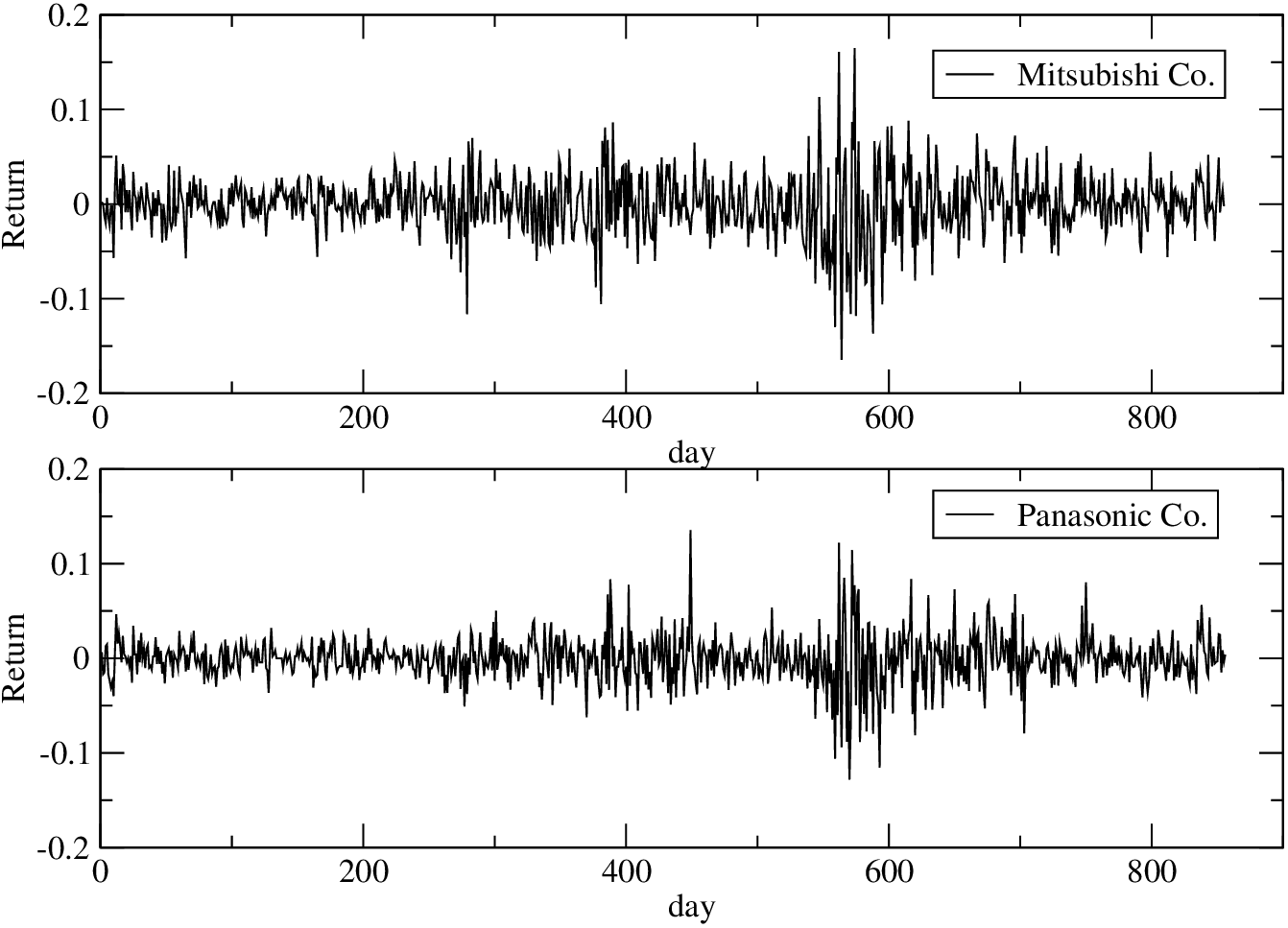}
\hspace{2mm}
\caption{
Daily close-to-close return time series of Mitsubishi Co. (top) and Panasonic Co. (bottom).}
\label{fig:Rt}
\vspace{1mm}
\end{figure}

\begin{figure}
\vspace{2mm}
\centering
\includegraphics[height=7.0cm]{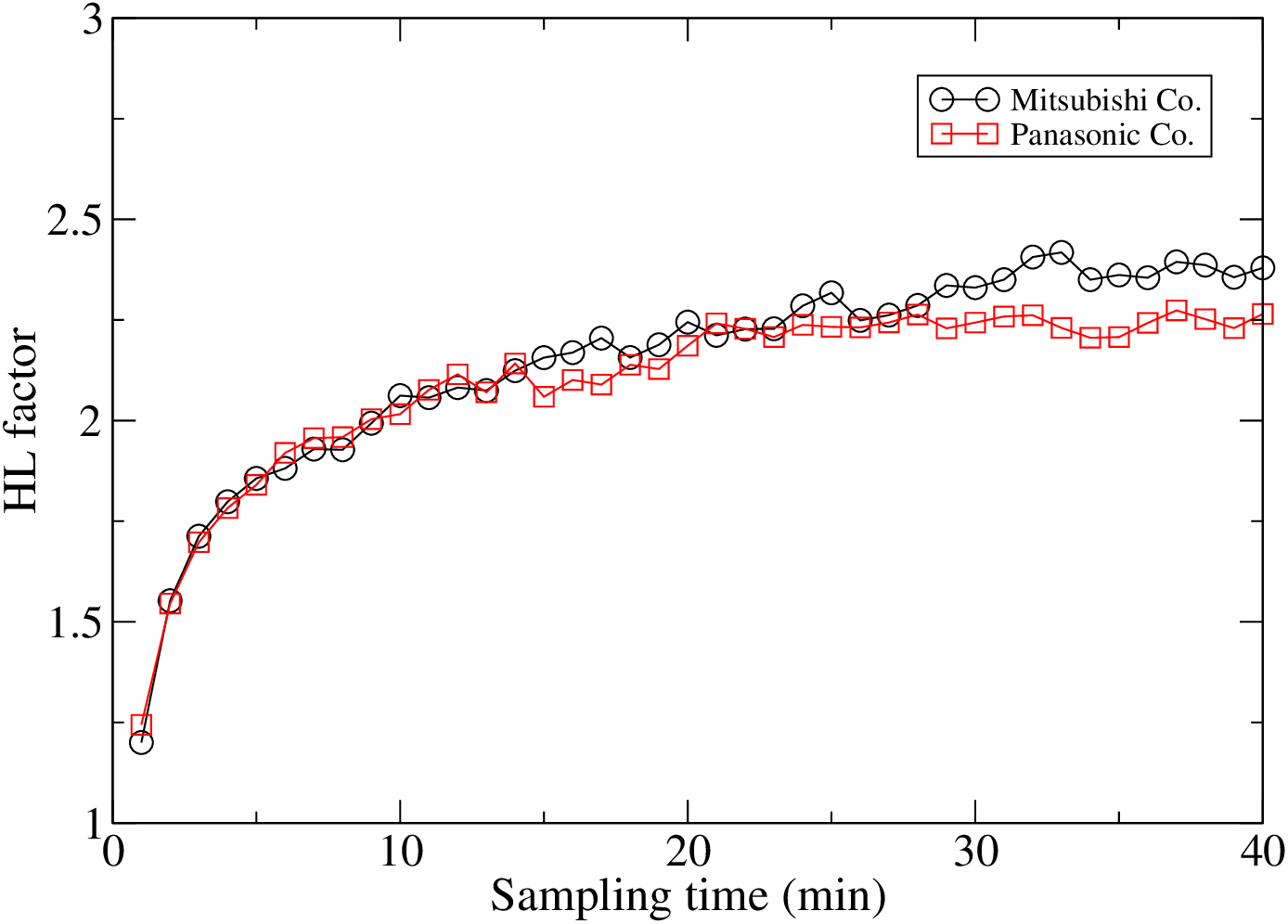}
\hspace{2mm}
\caption{
HL factor as a function of sampling time interval.
}
\label{fig:HL}
\end{figure}

\section{Empirical Analysis}
In this section we empirically study the SV model 
by applying it for asset returns traded on the Tokyo Stock Exchange.
The empirical study is based on daily data of two stocks (Mitsubishi Co. and Panasonic Co.).
These stocks are included in  the list of the Topix core 30 index which collects liquid stocks of the Tokyo Stock 
Exchange.
The sampling period of the stock data is  June 3, 1996 to December 30, 2009.
Figure \ref{fig:Rt} shows the daily close-to-close returns of the two stocks.

Using the daily close-to-close return data of Mitsubishi Co. and Panasonic Co.
we perform the Bayesian inference by the MCMC method for the SV model.
The first 10000 Monte Carlo samples are discarded and then 40000 samples are recorded for the analysis. 
The MCMC sampling of the volatility variables is performed by the HMC method.
The MCMC sampling of the model parameters $(\mu, \phi,\sigma^2_\eta)$ is implemented 
by the standard MCMC method\cite{SV}.
The values of the model parameters obtained by the MCMC method are summarized in table 1. 
We also list the results of $h_{10}$ in the table as a representative one of volatility variables.
It is found that the autocorrelation time of $h_{10}$ is not big but around 30 
which is much smaller than that of the  Metropolis algorithm\cite{SVHMC}. 

We measure the accuracy of the volatility from the SV model 
by calculating the following loss function.
\be
RMSPE= \sqrt{\frac1N \sum_{t=1}^N \left(\frac{{\bar{\sigma}_t^2} -cRV_t}{cRV_t}\right)^2},
\ee
where RMSPE stands for the Root Mean Squared Percentage Error and
$\bar{\sigma}_t^2$ is a volatility value at time $t$ averaged over 
the volatility data sampled by the HMC.

The realized volatility is calculated using sampling time from 1-min. to 20-min.
For each sampling time we also calculate the HL factor of eq.(\ref{eq:HL}).
Figure \ref{fig:HL} shows the HL factors of Mitsubishi Co. and Panasonic Co as a function of sampling time interval.
We find that both the HL factors from two stocks behave similarly and 
increase with sampling time interval. 
The smaller HL factors at small sampling time interval, i.e. at high sampling frequency are caused 
by the microstructure noise which artificially inflates the realized volatility.
At bigger sampling time interval, i.e. at lower sampling frequency where the microstructure noise effect is 
negligible the HL factors take bigger than 2, which means that 
returns during non-trading hours moves as much as returns during trading hours. 
Especially it is found that overnight return change dominates in non-trading hours\cite{TakaishiRV}. 

\begin{figure}
\centering
\includegraphics[height=7.0cm]{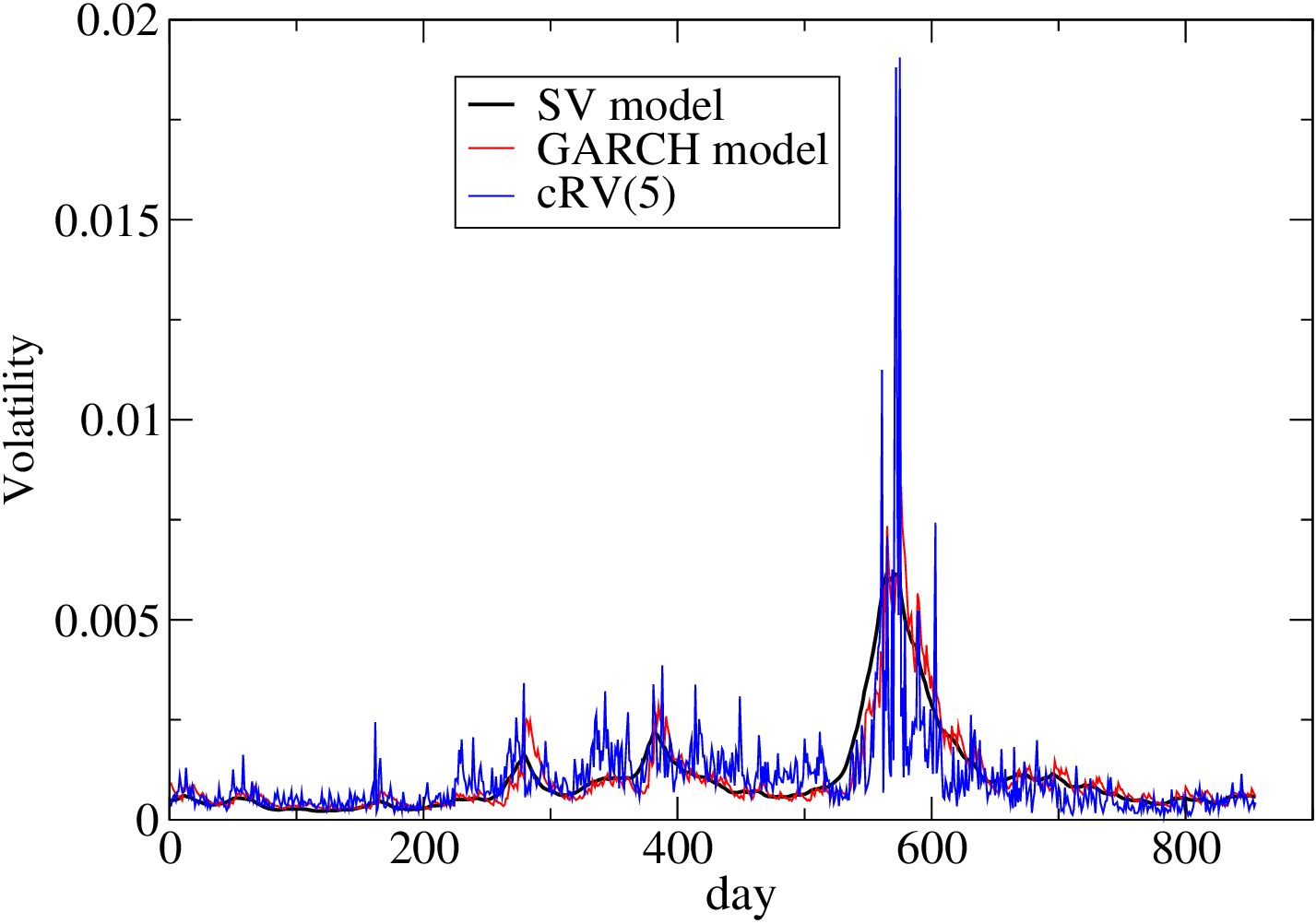}
\hspace{2mm}
\caption{
Volatility of Mitsubishi Co. obtained from the SV model, the GARCH model and the realized volatility ($cRV(5)$)
at 5-min sampling frequency. }
\label{fig:Vol1}
\vspace{3mm}
\end{figure}

\begin{figure}
\vspace{5mm}
\centering
\includegraphics[height=7.0cm]{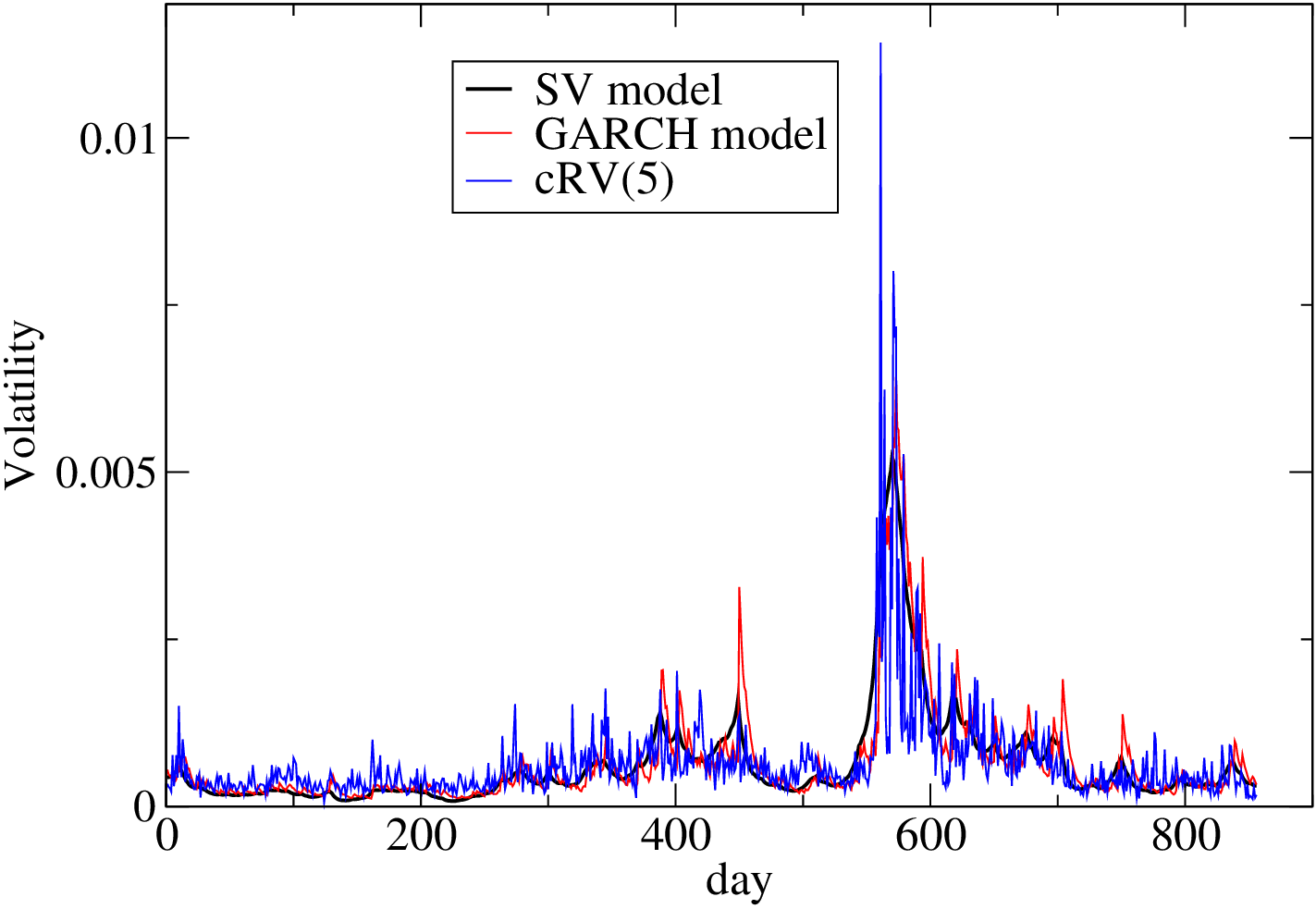}
\hspace{2mm}
\caption{
Volatility of Panasonic Co. obtained from the SV model, the GARCH model and the realized volatility ($cRV(5)$)
at 5-min sampling frequency.
}
\label{fig:Vol2}
\end{figure}

Using the HL factor $c$ we scale the realized volatility $RV_t$ to $cRV_t$.
Figure \ref{fig:Vol1}-\ref{fig:Vol2} compare the scaled realized volatility $cRV_t$ and the volatility from the SV model.
In the figures we only show the realized volatility constructed at 5-min sampling frequency 
as a representative one. 
The volatility estimated from the GARCH model is also shown in the figures.
We used the GARCH(1,1) model with normal errors which
is commonly used in empirical analysis of asset volatility.
The MCMC method\cite{Takaishi1,Takaishi2,Takaishi3,Takaishi4} 
was also used for the volatility estimation of the GARCH model.

\begin{figure}
\vspace{2mm}
\centering
\includegraphics[height=5.2cm]{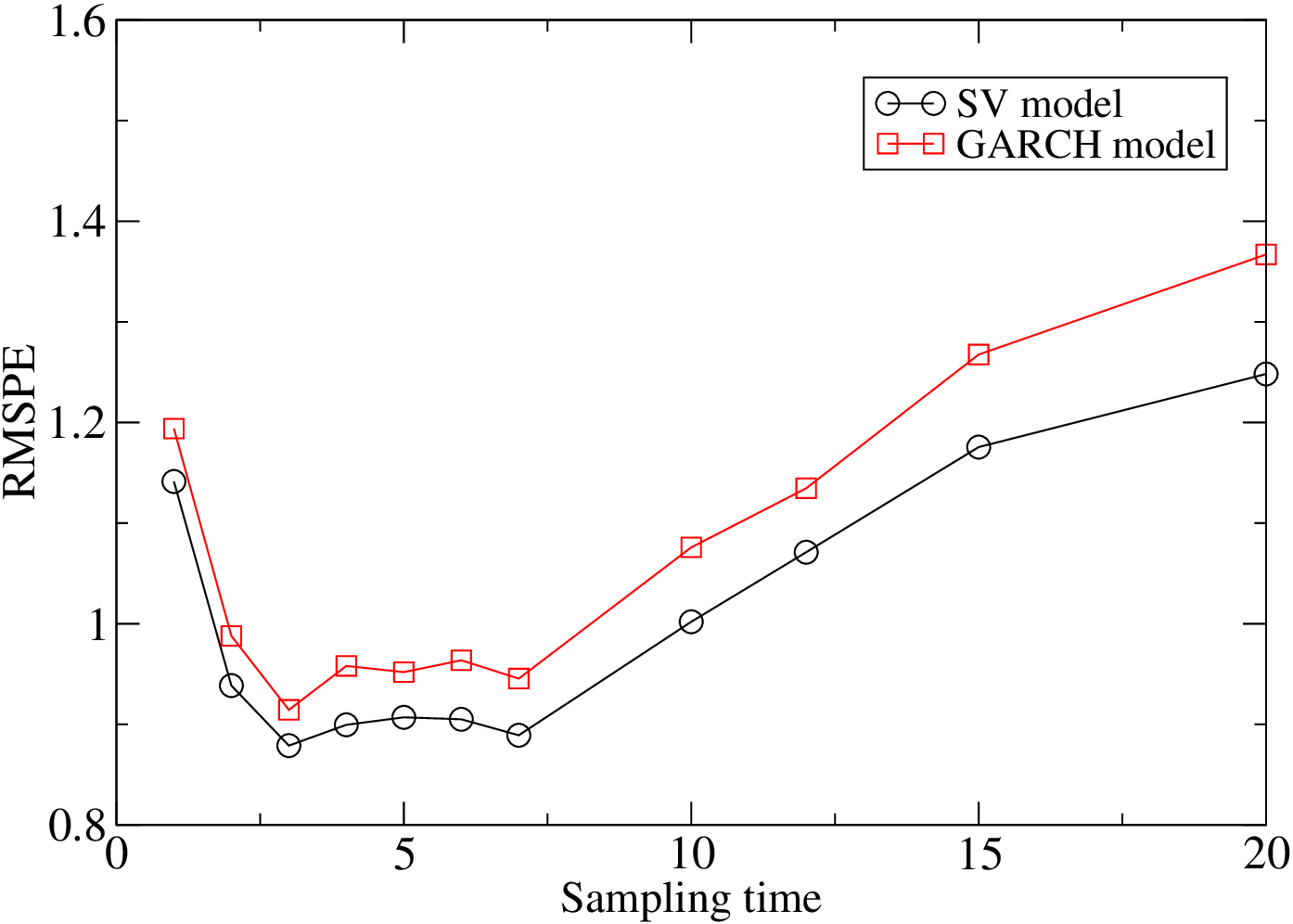}
\hspace{2mm}
\includegraphics[height=5.1cm]{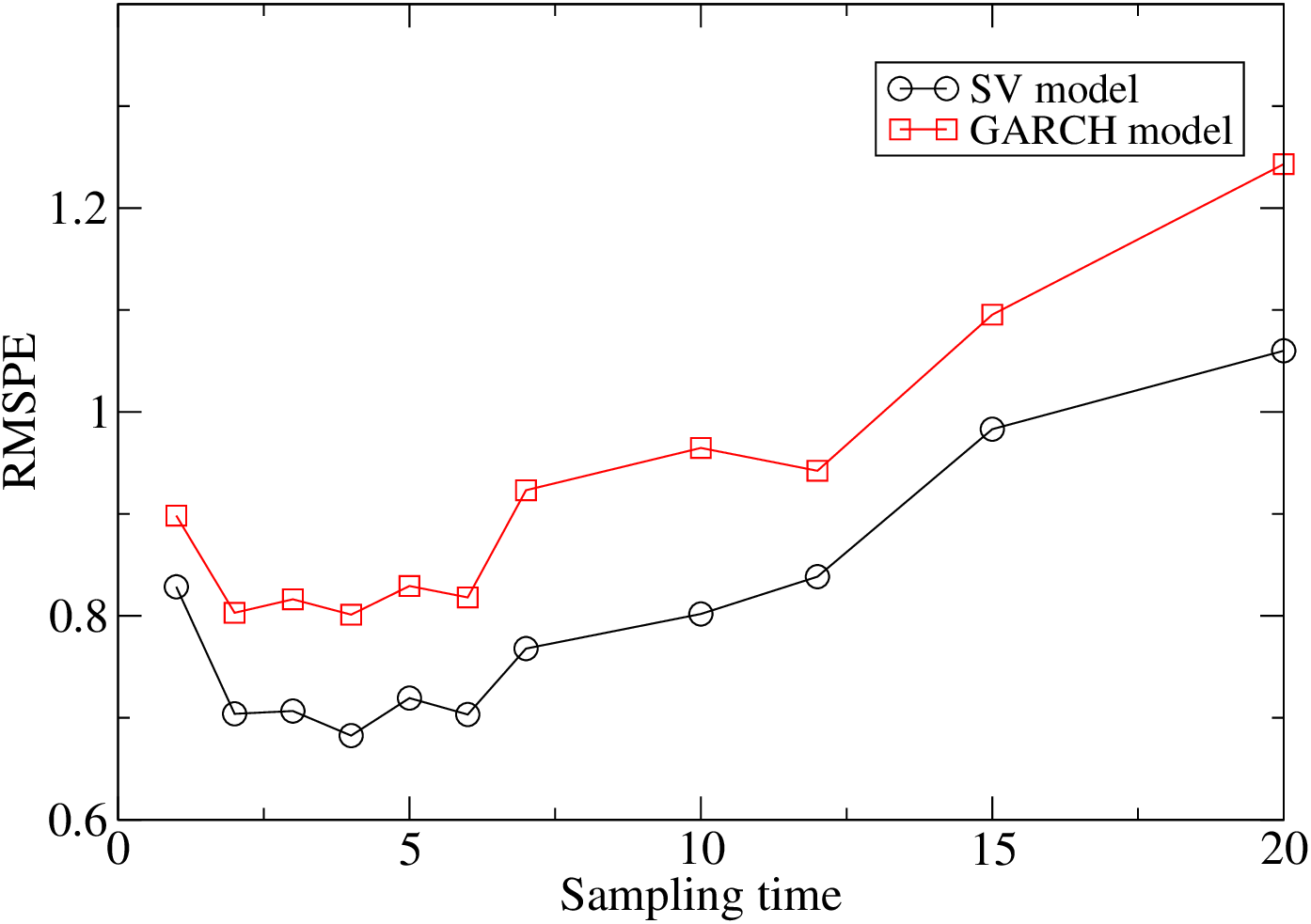}
\hspace{2mm}
\caption{
RMSPE of the SV model and the GARCH model for Mitsubishi Co. (left) and Panasonic Co. (right).}
\label{fig:RMSPE}
\vspace{-1mm}
\end{figure}

Figure \ref{fig:RMSPE} shows the RMSPE of the SV and the GARCH model.
It is seen that the RMSPE of the SV model is smaller than that of the GARCH model
which indicates that the SV model is superior to the GARCH model.
It is also interesting to notice that the minimum of the RMSPE takes around 5-min and
this observation is consistent with the optimum sampling time interval suggested in \cite{Bandi}.

\section{Conclusions}
We applied   the HMC algorithm to the Bayesian inference of the SV model
and estimated the volatility of two stock returns (Mitsubishi Co. and Panasonic Co.) 
traded on the Tokyo Stock Exchange.
We find that the volatility variables sampled by the HMC algorithm are well decorrelated  
compared to the Metopolis algorithm.
In order to quantify the accuracy of the estimated volatility
we calculated the RMSPE using 
the realized volatility modified by the HL factor as a proxy of the true volatility.
Comparing the RMSPE of the SV model with that of the GARCH model
we find that the volatility accuracy of the SV model is superior to that of the GARCH model.
Therefore empirically the SV model is preferred to the GARCH model 
based on the criterion of the volatility accuracy.

Since in this study we only used the GARCH(1,1) model it might be interesting  further to compare 
the SV model with other relevant GARCH-type models.

\subsection*{Acknowledgments.}
Numerical calculations in this work were carried out at the
Yukawa Institute Computer Facility
and the facilities of the Institute of Statistical Mathematics.
This work was supported by Grant-in-Aid for Scientific Research (C) (No.22500267).

\end{document}